\documentclass[a4paper, oneside, twocolumn, notitlepage, 10pt]{extarticle_ecoc}
\usepackage{ecoc}
\usepackage{tw}

\tikzset{
d0/.style={
    TUMred,
    line width=1pt,
    mark=*, 
    mark size=1pt,
},
d1/.style={
    TUMblue,
    line width=1pt,
    mark=*, 
    mark size=1pt,
},
dopt/.style={
    black,
    line width=1pt,
    mark=none, 
    mark size=1pt,
},
dPAM/.style={
    TUMOrange,
    line width=0.6pt,
    mark=none, 
    mark size=1pt,
},
dASK1/.style={
    TUMGreen,
    line width=0.6pt,
    mark=none, 
    mark size=1pt,
},
dASK2/.style={
    parula3,
    line width=0.6pt,
    mark=none, 
    mark size=1pt,
},
6ASK/.style={
    purple,
    line width=0.6pt,
    mark=triangle*, 
    mark size=2pt,
},
}

\begin{document}
\selectlanguage{english}    

\title{Optimizing Bipolar Constellations for High-Rate Transmission in Short-Reach Fiber Links with Direct Detection}


\author{
    Thomas Wiegart\textsuperscript{(1)}, Daniel Plabst\textsuperscript{(1)}, Norbert Hanik\textsuperscript{(1)}, Gerhard Kramer\textsuperscript{(1)}
}

\maketitle                  


\begin{strip}
    \begin{author_descr}
    
        \textsuperscript{(1)} Institute for Communications Engineering, School of Computation, Information and Technology, Technical University of Munich, Germany,
        \textcolor{blue}{\uline{thomas.wiegart@tum.de}}.
    \end{author_descr}
\end{strip}

\renewcommand\footnotemark{}
\renewcommand\footnoterule{}


\begin{strip}
    \begin{ecoc_abstract}
        Bipolar modulation increases the achievable information rate of communication links with direct-detection receivers. This paper optimizes bipolar transmission with a modulator bias offset for short-reach fiber links. A neural network equalizer with successive interference cancellation is shown to gain over 100~Gbit/s compared to standard receivers. 
        \textcopyright2024 The Author(s)
    \end{ecoc_abstract}
\end{strip}

\section{Introduction}
\Ac{DD} receivers are prominent in short-reach fiber-optic systems due to their low complexity, low energy consumption, and low cost compared to coherent receivers \cite{chagnon_optical_comms_short_reach_2019,zhong_dsp_for_short_reach_2018,qian_imdd_beyond_bw_lim_dcoi_2019}. \Ac{DD} is usually paired with unipolar intensity modulation, e.g., on-off keying. However, if the dominant noise is added before the receiver, \ac{DD} with oversampling has \acp{AIR} within $1$~\ac{bpcu} of the capacity of a coherent system \cite{mecozzi18_information,tasbihi20_capacity}. The paper \cite{plabst22_achievable} calculates \acp{AIR} for bipolar and complex signaling and performs \ac{JDD}.
Experiments demonstrating the gains of bipolar signaling are provided in \cite{wiegart22_experiments}. To reduce \ac{JDD} complexity, the paper \cite{prinz2024successive} proposes \ac{SIC}. Recently, a \ac{NN}-based \ac{SIC} receiver was proposed \cite{plabst24_neural} that can track large channel memory with low complexity. 

This paper evaluates the performance of the \ac{NN} receiver in \cite{plabst24_neural} for channels with bandwidth limitations and \ac{CD}. Our focus is optimizing the modulator bias offset. Note that, for a sufficiently large offset, a bipolar modulation becomes a unipolar modulation.
We show that optimizing the offset significantly increases \acp{AIR} under electrical bandwidth limitations compared to standard unipolar or bipolar modulations. We further show that the NN receiver compensates \ac{CD} through tracking the memory, and that \ac{SIC} provides large \ac{AIR} gains at high rates. For example, one can transmit in excess of $\SI{400}{Gbit/s}$ over a \ac{SSMF} of length $\SI{10}{km}$ in the O-Band at intermediate-SNRs.

\section{System Model}
\begin{figure*}
    \centering
    \includegraphics{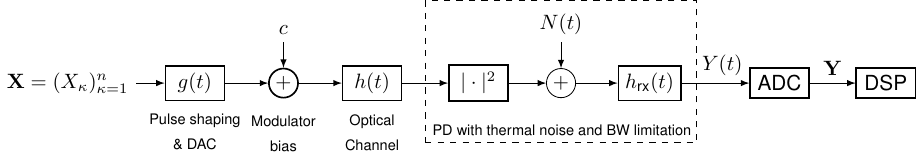}
    \caption{System model.\vspace{0.5cm}}
    \label{fig:systemmodel}
\end{figure*}

Fig.~\ref{fig:systemmodel} shows the continuous time baseband system model. Consider symbols $(X_\kappa)_{\kappa =1}^n$ drawn independent and identically distributed (iid) from a non-offset alphabet $\cX$ that we choose as amplitude shift keying (ASK), e.g., $\cX_{8\text{-ASK}} = \{\pm 1, \pm 3, \pm 5, \pm 7\}$. The symbols are pulse-shaped and offset by $c \ge 0$. In practice, a zero-mean pulse-shaped signal is generated by a \ac{DAC}, and the offset $c$ is obtained by tuning the modulator bias. We obtain the transmit signal
\begin{equation}
    X(t) = \sum_{\kappa=1}^n X_\kappa g(t-\kappa\Tsym) + c
\end{equation}
where $\Tsym$ is the symbol time and $\Rsym= 1/\Tsym$ is the symbol rate.

The signal $X(t)$ is transmitted into the fiber channel with linear response $h(t)$ and received by a \ac{PD} that outputs a current proportional to the absolute value squared of the impinging optical signal. We assume the \ac{PD} and the electrical components afterwards (trans-impedance amplifier and \ac{ADC}) are the dominant noise sources and also the components with the dominant bandwidth limitations. The received signal is
\begin{equation}
    Y(t) = h_\text{rx}(t) *  \left( \left| X(t) * h(t) \right|^2 + N(t)\right)
\end{equation}
where $h_\text{rx}(t)$ models the receiver bandwidth limitation and $N(t)$ is \ac{AWGN}.
The ADC samples at the Nyquist rate of $Y(t)$ and the samples of the channel outputs are collected in $\mathbf{Y}$. 

\section{Neural Network Equalizer}
We use the \ac{NN} receiver in~\cite{plabst24_neural} that is designed for multi-level coding with \ac{SIC}~\cite{PfisterAIRFiniteStateChan2001,wachsmann_multilevel_1999,prinz2024successive} and can approach \ac{JDD} \acp{AIR}. The \ac{NN} mimics the forward-backward algorithm~\cite{bcjr_1974} and outperforms existing \ac{JDD} receivers with substantially lower complexity. Moreover, the \ac{NN} exploits \ac{ISI} for phase-retrieval after \ac{DD}, making it compatible with power-efficient bipolar or complex-valued modulations; see~\cite{plabst22_achievable,tasbihi21,secondini20}. 

Let $I_S$ be the \ac{AIR} for $S$ \ac{SIC} levels in \ac{bpcu}, e.g., \ac{SDD} has $S=1$ and the \ac{AIR} $I_\text{SDD}=I_1$ \ac{bpcu}. We have
\begin{align}
    I_S \leq \lim_{n \rightarrow \infty }\frac{1}{n}  I(\mathbf{X}; \mathbf{Y}) \quad \mathrm{[bpcu]}.
    \label{eq:rate}
\end{align}
where the right-hand side of \eqref{eq:rate} is the \ac{JDD} \ac{AIR}.
The net bit rate is
\begin{align}
    R_\text{b} = I_S \cdot \Rsym \quad \mathrm{[bit/s]}.  
\end{align}
A moderate number of SIC stages, e.g., $S=2,\dots, 8$ often give large \ac{AIR} gains; see~\cite{prinz2024successive, plabst24_neural,jaeger2024information}.

\section{System Parameters and Offset Optimization}
Consider the following system parameters: at the transmitter, we use a frequency-domain raised-cosine pulse with roll-off factor $\alpha=0.15$. The symbol alphabet $cX$ is scaled to the interval $[-1, 1]$ before pulse shaping, e.g., $c=0$ and $c=1$ corresponds to pure bipolar and unipolar signaling, respectively. (This does not preclude negative overshoots of the continuous-time signal. Note that a realistic system with a \ac{MZM} operated at the quadrature point usually has $c \gg 1$.)

We study transmission over $\SI{10}{km}$ of \ac{SSMF} operated in the O-Band (\ac{CD} with $\beta_2 = \SI{-2}{ps^2/km}$ and dispersion-slope $\beta_3 = \SI{0.07}{ps^3/km}$). We assume negligible fiber nonlinearity. At the receiver, we consider a p-i-n \ac{PD} at the thermal noise limit and calculate the noise power for a given \ac{ROP} according to Eq.~(4.4.13) in\cite{agrawal} (with  $R=\SI{0.9}{A/W}$, $4k_BTF_n/R_L=\SI{3e-22}{A^2s}$, and $\Delta f = \SI{100}{GHz}$). The $\SI{3}{dB}$ cutoff frequency of $h_\text{rx}(t)$ is $\SI{95}{GHz}$.
The received signal is re-sampled to two \ac{SPS} and fed to the \ac{NN} receiver \cite{plabst24_neural}. The NN has an input layer with $64$ neurons and two hidden layers with $256$ neurons each, followed by an output layer with $|\cX|$ outputs. The NN performs \ac{SIC} with $S=3$ stages.

\section{Offset Optimization}
\begin{figure*}
    \fbox{
    \parbox{.975\linewidth}
    {\footnotesize{\hfill
    \pltref{d0} $c=0$ (diff. enc)\hfill
    \pltref{d1} $c=1$ (w/o diff. enc)\hfill
    \pltref{dopt} opt. $c$ (w/o diff. enc)\hfill
    \pltref{6ASK} $6$-ASK (opt. $c$ w/o diff. enc)
    \hfill\phantom{.} \\
    {\phantom{.}}\hfill
    all $c$ w/o diff. enc: \hfill
    \pltref{dASK1} $c \in \{0.0,0.1,0.2,0.3,0.4\}$\hfill
    \pltref{dASK2} $c \in \{0.5,0.6,\dots, 0.9\}$\hfill
    \pltref{dPAM} $c \in \{1.1, 1.2, \dots, 1.5, 2.0\}$
    \hfill\phantom{.}
    }}
}
\vspace{0.1cm}
    \begin{subfigure}[t]{0.35\textwidth}
        \centering
        \includegraphics{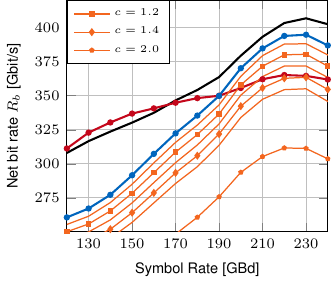} \\
        \includegraphics{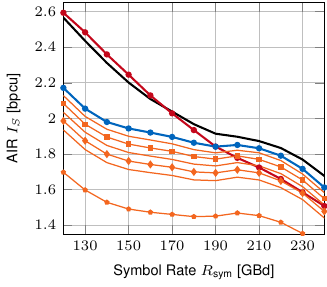}        
        \caption{Baseline curves ($c=0$, $c=1$, opt. $c$ (w/o diff. enc) and offsets $c>1$ (orange curves).}
    \end{subfigure}
    \begin{subfigure}[t]{0.32\textwidth}
        \includegraphics{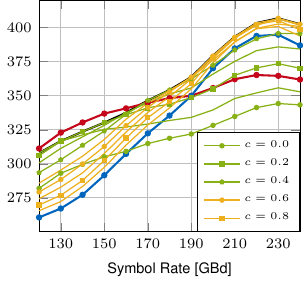} \\
        \includegraphics{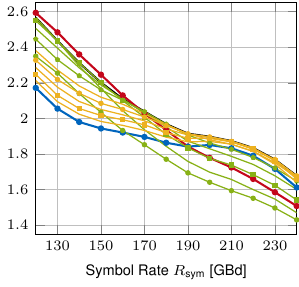}
        \caption{Baseline curves and offsets (w/o diff. enc) $c \in [0.0, 0.4]$ (green curves) and $c \in [0.5, 0.9]$ (yellow curves).}
    \end{subfigure}
    \begin{subfigure}[t]{0.32\textwidth}
        \includegraphics{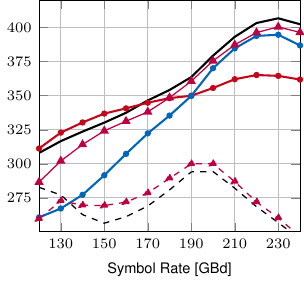}\\
        \includegraphics{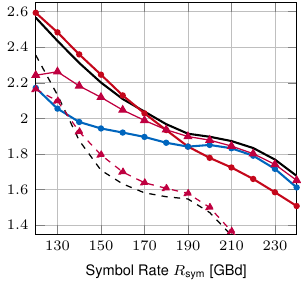}
        \caption{Baseline curves and comparison to SDD without SIC (dashed curves), as well as to $6$-ASK with $S=3$ and optimized offset.}
    \end{subfigure}
    \vspace{0.3cm}
    \caption{Net bit rates $R_b$ (top figures) and AIRs $I_S$ (bottom figures) for different offset values with $8$-ary modulation, transmission over $\SI{10}{km}$ of SSMF in the O-Band, ROP of $\SI{-15}{dBm}$ and $S=3$ SIC stages.\vspace{0.7cm}} 
    \label{fig:offset7dB}
\end{figure*}

Fig.~\ref{fig:offset7dB} shows the results of the offset optimization at a \ac{ROP} of $\SI{-15}{dBm}$. The top subfigures show net bit rates $R_b$ for various symbol rates and offsets while the bottom figures show the corresponding \acp{AIR} $I_s$. All subfigures show curves for:
\smallskip
\begin{itemize}
    \item $c=0$ (pure bipolar signaling) in red;
    \item $c=1$ (unipolar signaling) in blue;
    \item the optimal offset $c$ in black.
\end{itemize}
For $c=0$, we used differential precoding as in \cite{plabst22_achievable, plabst24_neural}. For $c>0$, we did not use differential precoding because the \ac{NN} receiver could recover the transmit signal for the best $c$, except for minor losses at low symbol rates. The left subfigures show the performance for $c>1$. This corresponds to a system with a \ac{MZM} biased at the quadrature point. The performance decreases with increasing offset $c$.

The middle subfigures show the performance for $c \in [0.0, 0.4]$ (green curves) and $c \in [0.5, 0.9]$ (yellow curves). For small $c$, the performance suffers mainly due to the lack of differential precoding; compare the green curve with markers ($c=0$) with the red baseline curve. Moderate $c$ values (yellow curves) bridge the gap between $c=0$ and $c=1$ at intermediate symbol rates and outperform the peak of the $c=1$ curve at high symbol rates. A peak bit rate of $R_\text{b} = \SI{407}{Gbit/s}$ is obtained with $c=0.6$ at a symbolrate of $R_\text{sym}=\SI{230}{GBaud}$.

The right subfigures compare the performance of the NN receiver with $S=3$ SIC stages (as in the other figures) with an SDD receiver, i.e., $S=1$ (dashed curves). The results highlight the significant performance gains via SIC in bandlimited scenarios. Observe that classic SDD receivers achieve less than $\SI{300}{Gbit/s}$ (even with optimized modulator bias; see the black dashed curve), and SIC increases the peak net bit rate by over $\SI{100}{Gbit/s}$. We also plot the performance of $6$-ASK with $S=3$ (purple curve with triangles).
With an optimal $c$, the $6$-ASK performance is only slightly worse, while allowing for a lower FEC overhead. Under SDD, $6$-ary signaling slightly outperforms $8$-ary signaling (dashed curves).

\begin{figure*}
    \begin{subfigure}[t]{0.35\textwidth}
        \centering
        \includegraphics{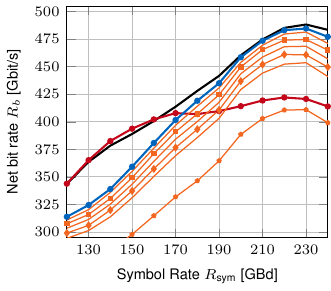} \\
        \includegraphics{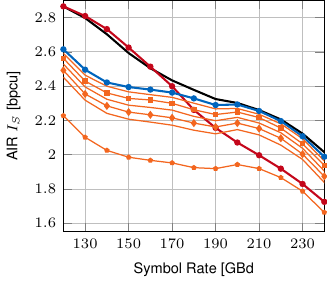}     
        \caption{Baseline curves ($c=0$, $c=1$, opt. $c$ (w/o diff. enc) and offsets $c>1$ (orange curves).}
    \end{subfigure}
    \begin{subfigure}[t]{0.32\textwidth}
        \includegraphics{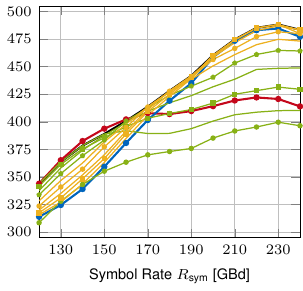} \\
        \includegraphics{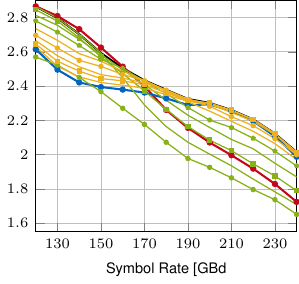}
        \caption{Baseline curves and offsets (w/o diff. enc) $c \in [0.0, 0.4]$ (green curves) and $c \in [0.5, 0.9]$ (yellow curves).}
    \end{subfigure}
    \begin{subfigure}[t]{0.32\textwidth}
        \includegraphics{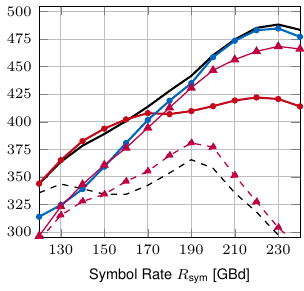}\\
        \includegraphics{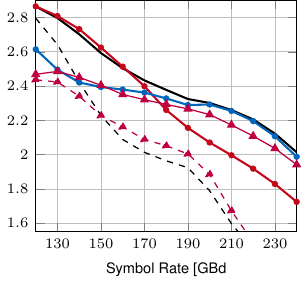}
        \caption{Baseline curves and comparison to SDD without SIC (dashed curves), as well as to $6$-ASK with $S=3$ and optimized offset.}
    \end{subfigure}
    \vspace{0.3cm}
    \caption{Net bit rates $R_b$ (top figures) and AIRs $I_S$ (bottom figures) for different offset values with $8$-ary modulation, transmission over $\SI{10}{km}$ of SSMF in the O-Band, ROP of $\SI{-13}{dBm}$ and $S=3$ SIC stages. See Fig.~\ref{fig:offset7dB} for the legend.}
    \label{fig:offset9dB}
\end{figure*}

Fig.~\ref{fig:offset9dB} shows the same curves for an ROP of $\SI{-13}{dBm}$ with similar effects as described above. As we are less noise-limited, the overall achievable net bit rate is higher, and the gains by optimizing $c$ are reduced, and also, the loss through $6$-ASK is slightly higher.

\section{Conclusions}
We studied the performance of bipolar transmission with an optimized modulator offset for \ac{DD} receivers. The NN-SIC equalizer from \cite{plabst24_neural} achieves a net bit rate of over $\SI{400}{Gbit/s}$ for a $\SI{10}{km}$ link operated in the O-Band (ROP $\SI{-15}{dBm}$, receiver bandwidth $\SI{95}{GHz}$). Furthermore, SIC gains over $\SI{100}{Gbit/s}$ compared to standard \ac{SDD} receivers used in current systems. Future work is planned to experimentally demonstrate the simulated results.

\clearpage


\printbibliography

\vspace{-4mm}

\end{document}